\documentclass[a4,11pt]{article}
\usepackage[fleqn]{amsmath}
\usepackage{amsthm}
\usepackage{newtxtext}
\usepackage[varg]{newtxmath}
\usepackage[noend]{algpseudocode}
\usepackage{algorithm}
\usepackage{braket}
\usepackage{url}

\newcommand*{\refAlgo}[1]{{Algorithm~\ref{#1}}}
\title{Maximum Number of Steps of Topswops on $18$ and $19$ Cards}
\author{Kento Kimura, Atsuki Takahashi, Tetsuya Araki, Kazuyuki Amano\footnote{All authors are at Department of Computer Science, Gunma University}}

\begin{document}
\maketitle

\begin{abstract}
Let $f(n)$ be the maximum number of steps of {\it Topswops} on $n$ cards.
In this note, we report our computational experiments to
determine the values of $f(18)$ and $f(19)$.
By applying an algorithm developed by Knuth in a parallel fashion,
we conclude that $f(18)=191$ and $f(19)=221$.

\end{abstract}
\section{Introduction}
Consider a deck of $n$ cards numbered $1$ to $n$ arranged in random order, which
can be viewed as a permutation on $\{1,2,\ldots, n\}$.
Continue the following operation until the top card is $1$.
If the top card of the deck is $k$,
%count $k$ cards from the top of the deck and turn over the block of the cards.
then turn over a block of $k$ cards at the top of the deck.
This card game is called \textit{Topswops}, which was originally invented by J.H. Conway in 1973. See e.g., the introduction of \cite{TheoreticalLB} for a short history of the game.

The problem is to find an initial deck that requires a maximum number of steps until termination, for a given number of cards.
%A deck of cards can be viewed as a permutation of $\Set{1,2,\cdots,n}$.
%A starting deck is called \textit{initial}.
For a positive integer $n$,
let $f(n)$ be the maximum number of steps until termination for Topswops on $n$ cards.
We call an initial deck that needs $f(n)$ steps \textit{largest}.
For example, the deck $(3,1,4,5,2)$ is largest for $n=5$. 
The game goes as
%\begin{quote}
\begin{eqnarray}
\label{Eq:example}
(\underline{3,1,4},5,2)  \rightarrow   (\underline{4,1,3,5},2)
\rightarrow (\underline{5,3,1,4,2}) \rightarrow (\underline{2,4},1,3,5) \qquad \nonumber \\
\rightarrow (\underline{4,2,1,3},5) \rightarrow 
(\underline{3,1,2},4,5) \rightarrow (\underline{2,1},3,4,5)
\rightarrow (1,2,3,4,5),
%\end{quote}
\end{eqnarray}
and terminates after $f(5)=7$ steps.

The best known upper bound on $f(n)$ is $F(n+1)-1=O(1.618^n)$, where $F(k)$ is the $k$-th Fibonacchi number \cite[Problems 107--109]{Knuth} and
the best known lower bound is $\Omega(n^2)$ \cite{TheoreticalLB}.
The gap is exponential.
The exact values of $f(n)$ for $n \leq 17$ have been obtained by an exhaustive search with some
pruning techniques. %(see \cite[A000375]{OEIS} and the references therein). 
The sequence is $(0, 1, 2, 4, 7, 10, 16, 22, 30, 38, 51, 65, 80, 101, \linebreak 113, 139, 159 )$ for $n=1,2,\ldots, 17$.
See the sequence A000375 of OEIS \cite{OEIS}.

In this note, we describe our effort for extending this list for $n=18$ and $19$.
Namely, by applying an algorithm developed 
by Knuth \cite{KnuthCode} in a parallel fashion, we conclude that $f(18)=191$ and $f(19)=221$. 
We also find that the number of initial decks that attain the maximum for $n=18$ is one and that for $n=19$ is four, respectively.
Note that the values for $n \leq 17$ are listed as the sequence A123398 at OEIS \cite{OEIS}.

The rest of this note is as follows.
In Section 2, we give a brief explanation of Knuth's algorithm 
\cite{Knuth}. %(\cite[Solution of Problem 107 on page 119]{Knuth}). 
Then, in Section 3,
we describe our computational experiments for determining $f(18)$ and $f(19)$.
The code used in our experiments can be viewed on GitLab
at \url{https://gitlab.com/kkimura/tswops}.

\section{Knuth's algorithm}
In this section, we explain an algorithm for finding a largest deck for Topswops used in our experiment, which was developed by Knuth \cite[Solution of Problem 107]{Knuth} (see also \cite{KnuthCode} for the code itself).
Three algorithms were described there and we use the most efficient one, which is referred to as a ``better" algorithm.

%For a positive integer $n$, $[n]$ denotes the set $\Set{1,2,\cdots,n}$.
%A deck of cards for the game can be viewed as a permtation on $[n]$.
For a natural number $n$,
let $[n]$ denote the set $\{1,2,\ldots, n\}$.
For an initial deck $A$, let $S(A)$ be a list $(d_1, d_2, \ldots, d_k)$ ($k \leq n$) where $d_i$ is the $i$-th card that appeared at the top of the deck in the game starting from $A$. 
For example, $S((3,1,4,5,2))=(3,4,5,2,1)$ (see Eq. (\ref{Eq:example}))
and $S((3,5,4,1,2))=(3,4,1)$. 
Notice that the length of $S(A)$ depends on $A$, but the
last element of $S(A)$ is always $1$.
An important property is that if $A$ is largest, then the length of $S(A)$ must be $n$.
This can be verified by seeing that if $S(A)=(d_1, \ldots, d_{k-1}, 1)$ for some
$k < n$, then we can always create another deck $A'$ such that the first $k$ elements of $S(A')$ is $(d_1, \ldots, d_{k-1}, d')$ for $d' \in \{2,\ldots, n\} \backslash \{d_1, \ldots, d_{k-1}\}$ and that the game for $A'$ is strictly longer than the one for $A$.

Let $P$ be the set of all lists $p=(p_1,p_2, \ldots, p_n)$
such that $p$ is a permutation on $[n]$ and $p_n=1$.
Given a list $p \in P$, we can get an initial deck $S^{-1}(p)$ by the following algorithm.
In \refAlgo{algorithm:tryswops},
the minus value $-i$ in $A$ means
that the $i$-th card in a deck is not specified yet.
%For a permutation $p$ on $[n]$, $p_i$ denotes the $i$-th element of $p$.
%Let $I$ denote the set of all permutations of $[n]$.
%For an initial deck $a\in I$,
%let $g(a)$ be the number of steps in a game starting with $a$.
%Let $P \subset I$ be the set of all permutations on $[n]$ such that $p_n=1$.
%It is obvious that $|P|=(n-1)!$.

%We introduction an algorithm that outputs an initial decks for a given sequence $p\in P$.
%See \refAlgo{algorithm:tryswops}.
\begin{algorithm}[htb]
    \caption{Generate an Initial Deck}
    \label{algorithm:tryswops}
    \begin{algorithmic}[1]
        \Procedure{GenInitDeck}{$p$}
            \State Let $A$ be an array with $(-1,-2,\ldots,-n)$.
            \For{$i=1,2,\ldots, n$}
                \State $a_{-A_{1}}\coloneqq p_{i}$ \label{algo:tryswops:SubstituteP}
                \State $A_{1}\coloneqq p_{i}$  \label{algo:tryswops:SubstituteP2}
                \While{$A_{1} > 1$}
                    \State Turn over a block of $A_{1}$ cards of $A$.
                \EndWhile
            \EndFor
            \State return $(a_{i})_{i\in[n]}$
        \EndProcedure
    \end{algorithmic}
\end{algorithm}

The above arguments suggest that we can determine $f(n)$ by examining all $(n-1)!$ lists in $P$ together with \refAlgo{algorithm:tryswops}.
Essentially, Knuth's algorithm enumerates these lists as well as corresponding decks in a depth-first fashion. 
Moreover, the algorithm applies two pruning criteria to reduce the size of  the search tree.

The first pruning is based on the fact that a largest deck must be a {\it derangement}, i.e.,
the $k$-th card from the top is not $k$ for every $k \in [n]$.
In order to explain the second pruning, we need some definitions.
Let $A$ be an initial deck and let $A_c$ be the deck obtained from $A$ by executing $c$ steps of the game. Let $T(A_c)$ denote the largest 
integer $k$ such that the cards numbered $1,2,\ldots, k$ are located
at positions at $1,2,\ldots, k$ (in an arbitrary order) in the deck $A_c$.
It is obvious that if $f(T(A_c)) + c < f(n)$, then $A$ is not largest.
Although $f(n)$ is not known beforehand, we can use any lower
bound $\ell(n)$ on  $f(n)$ in the right hand side of inequalities for pruning.

Note that the depth of the search tree without pruning is $(n-1)$
and  each node at depth $k$ has $n-1-k$ children.

\section{Experiments and Results}

Since the search tree of Knuth's ``better" algorithm is well-balanced,
it is easy to be parallelized.
First, we generate the search tree for the first few levels, which corresponds
to the first few elements of the list $p$ explained in the last section.
Then, distribute the leaves of the tree to many threads and resume the generation in parallel by letting a given leaf as a root of a subtree.

For $n=18$, we truncate the tree at level two and divide it into
$240$ subtrees.
For $n=19$, we truncate the tree at level three and divide it into
$3,952$ subtrees. Each of these numbers is slightly smaller than the
one in the original search tree, i.e., $272(=17 \times 16)$ or
$4,080(=18 \times 17 \times 16)$, because of the pruning.

In our experiments, we use up to 172 threads in parallel spreading out over nine standard PCs.
The computation takes about 7 hours for 
$n=18$ (using 132 threads), and about 6 days for $n=19$ (using 172 threads).
This means that, if we run the code on a single thread, then the computation would take approximately $10^3$ days for $n=19$.
The total numbers of traversed nodes are 
$43,235,268,208,065$ for $n=18$ and
$933,351,108,741,643$ for $n=19$, respectively.
The ratios to the number of nodes in the search tree without pruning,
i.e., $\sum_{i=0}^{n-1} \prod_{j=1}^{i} (n-j)$, are
$4.47\%$ and $5.36\%$, respectively.
The breakdown of the number of traversed nodes for $n=19$ with respect to the levels of the tree is shown in Table \ref{Tab:nodes}.

By examining the result, we conclude that $f(18)=191$ and $f(19)=221$.
The largest initial deck for $n=18$ is unique. It is
\begin{center}
(6 14 9 2 15 8 1 3 4 12 18 5 10 13 16 17 11 7),
\end{center}
which terminates at the sorted position (1 2 3 $\ldots$ 18).
There are four largest initial decks for $n=19$. These are
\begin{center}
(9 4 19 17 10 1 11 15 12 8 5 2 18 13 16 7 3 14 6),\\
(12 15 11 1 10 17 19 2 5 8 9 4 18 13 16 7 3 14 6),\\
(12 1 18 11 3 14 2 6 8 16 5 4 15 10 13 17 19 7 9),\\
(12 1 18 11 2 3 14 6 8 16 5 4 15 10 13 17 19 7 9).
\end{center}
 %[1, 10, 9, 8, 7, 6, 5, 4, 3, 2, 11, 12, 13, 14, 15, 16, 17, 18, 19]
 %[1, 10, 9, 8, 7, 6, 5, 4, 3, 2, 11, 12, 13, 14, 15, 16, 17, 18, 19]
 %[1, 10, 9, 8, 7, 6, 5, 4, 3, 2, 11, 12, 13, 14, 15, 16, 17, 18, 19]
 %[1, 10, 9, 8, 7, 6, 5, 4, 3, 2, 11, 12, 13, 14, 15, 16, 17, 18, 19]
Interestingly, all these decks terminate at a same non-sorted
position (1 10 9 8 7 6 5 4 3 2 11 12 13 14 15 16 17 18 19).
The largest initial deck that terminates at the sorted position
is known to take $209$ steps (see A000376 of OEIS \cite{OEIS}), which is twelve less than the value of $f(19)$. 
\begin{table} 
\caption{The number of traversed nodes for $n=19$.}
\label{Tab:nodes}
\begin{verbatim}
     Level   # of traversed nodes | Level   # of traversed nodes
        0                     1       10           46335514956
        1                    17       11          304773283939
        2                   272       12         1716889839183
        3                  3952       13         8059154346527
        4                 52861       14        30428256670076
        5                653126       15        89242470628183
        6               7419100       16       200111553921243
        7              77075852       17       326581145735086
        8             726678384       18       276853558861087
        9            6158057798   -----------------------------
       10           46335514956    Total       933351108741643
\end{verbatim} 
\end{table}
%       11          304773283939
%       12         1716889839183
%       13         8059154346527
%       14        30428256670076
%       15        89242470628183
%       16       200111553921243
%       17       326581145735086
%       18       276853558861087
\section*{Acknowledgements}
This work was partially supported by JSPS
Kakenhi Grant Numbers 18K11152 and 18H04090.

\end{document}